\documentclass[
reprint,
groupedaddress,
amsmath,amssymb,
prb,
floatfix,
superscriptaddress
]{revtex4-1}

\usepackage{amsmath}
\usepackage{bm}
\usepackage{amsfonts}
\usepackage{graphicx}
\usepackage{amssymb}
\usepackage{sidecap}
\usepackage[caption=false]{subfig}
\usepackage{hyperref}
\usepackage{cleveref}
\usepackage{siunitx}
\usepackage[T1]{fontenc}
\usepackage{libertine}
\begin{document}

\title{Investigating the electronic origins of the repulsion between substitutional and interstitial solutes in hcp Ti}
\author{N. S. Harsha Gunda}
\affiliation{Materials Department, University of California Santa Barbara}
\author{Carlos G. Levi}
\affiliation{Materials Department, University of California Santa Barbara}
\author{Anton Van der Ven}
\email{avdv@ucsb.edu}
\affiliation{Materials Department, University of California Santa Barbara}
\date{\today}

\begin{abstract}

The high solubility of oxygen in Ti, Zr and Hf makes it difficult to stabilize the protective oxide scales on their surfaces as the subsurface regions can serve as boundless sinks that continuously dissolve oxygen. 
Alloying elements are crucial to reduce the oxygen solubility and diffusivity within early transition metals. 
Past studies have shown that substitutional alloying additions to titanium repel interstitial oxygen. 
Here we use first-principles calculations to show that this repulsion is short ranged and identify a variety of factors that are likely responsible for the repulsive interaction. 
We identify a unique hybridization phenomenon between dissolved substitutional elements and interstitial oxygen within hcp Ti that leads to a repulsive interaction at short distances, similar to that between closed-shell atoms.
Calculations of Bader charges also suggest the existence of short-range Coulomb interactions due to the accumulation of charge on the substitutional solute and interstitial oxygen that is drawn from the Ti host. 

\end{abstract}

\maketitle
\section{Introduction}\label{sec:intro}

Early transition metals belonging to groups 4 and 5 of the periodic table (Ti, Zr, Hf, V, Nb and Ta) are able to dissolve high concentrations of interstitial species such as oxygen, nitrogen and carbon and form a rich family of oxides, nitrides and carbides  \cite{abriata1986zr,murray1987ti,okamoto2011ti,goldschmid2013interstitial,salamat2013nitrogen,paul2011first,burton2012TiO,burton2012HfO,puchala2013thermodynamics,chen2015high,gunda2018first,gunda2018resolving,weinberger2018review}. 
Ti, Zr, and Hf in particular can dissolve oxygen over the interstitial sites of their hcp crystal structure up to an atom fraction of 0.33.
This high solubility not only affects mechanical properties \cite{yu2015origin,ghazisaeidi2014interaction,chong2020mechanistic}, but makes it almost impossible to sustain a protective oxide scale on their surfaces as the scale is continuously undermined from below \cite{kofstad1961oxidation,kofstad1967high,bertrand1984morphology,unnam1986oxidation,chaze1987influence}. 
Alloying strategies are therefore essential to enable the stabilization of an oxide scale on metals such as titanium.
The alloying additions play several roles. 
They may alter the types of oxide phases that form during oxidation \cite{luthra1991stability,rahmel1991thermodynamic,kelkar1995phase,seifert2001phase}. 
They can also reduce the oxygen solubility and oxygen mobility in the base metal. \cite{chaze1987influence,wu2013solute,gunda2020understanding}

A first principles study by Wu and Trinkle \cite{wu2013solute} predicted that almost every element of the periodic table when substitutionally dissolved in hcp Ti will repel interstitial oxygen. 
The repulsion increases when going to the right in the periodic table.
Alloying elements such as Al and Si are predicted to have especially large and repulsive binding energies with interstitial oxygen that are on the order of 0.8 to 1 eV, respectively. 
This repulsion is surprising considering that elemental Al and Si readily react with oxygen to form Al$_2$O$_3$ and SiO$_2$.
A recent first-principles study of the Ti-Al-O ternary \cite{gunda2020understanding} confirmed the existence of a large repulsion between dissolved Al and oxygen in Ti and showed how this repulsion affects thermodynamic properties at elevated temperature.

Here we systematically investigate the origin of the repulsive interactions between interstitial oxygen and dissolved alloying elements within Ti using first-principles electronic structure calculations.  
We focus on a subset of the alloying elements considered by Wu and Trinkel and analyze the role of electronic structure, atomic relaxations and charge redistribution between host and solutes. 
Two distinct interactions are identified that have their origin in the unique electronic structure of a Ti host containing dilute interstitial oxygen. 
One interaction arises from the hybridization between atomic-like orbitals on dissolved oxygen and substitutional solutes that lead to a closed-shell type of repulsion at short distances. 
The second interaction is electrostatic in nature and is also short ranged, arising from a redistribution in charge between the Ti host and solutes. 
The insights generated by this study should also apply to oxygen-solute interactions in other early transition metals.

\section{Methods}

First-principles electronic structure calculations were performed using density functional theory as implemented in the VASP plane-wave software package.
\cite{kresse1993ab,kresse1994ab,kresse1996efficiency,kresse1996efficient} 
The exchange correlation functional was that of Perdew, Burke and Ernzerhof (PBE) \cite{perdew1996generalized}. 
The projector augmented wave method (PAW) was used to describe interactions between valence and core electrons.\cite{blochl1994projector,kresse1999ultrasoft}
The valence states of the Al and Si PAW-PBE potentials were the 3s and 3p states, while those of O were the 2s and 2p states. For Ti and Fe, belonging to the 3d transition element series, the "sv" potentials with valence states 3s, 3p, 3d and 4s states were used. 
For Cr, the "pv" potential with 3p, 3d and 4s valence states was used. 
The "sv" potentials with 4s, 4p, 4d and 5s valence state were used for Y, Zr and Nb in the 4d transition element series.
All calculations were performed with a plane wave cutoff of 550 eV. 

Calculations were performed using a $3\times3\times3$ super cell of hcp Ti with a $\Gamma$-centered $6\times6\times3$ k-point grid. 
The volume was held fixed at the equilibrium volume of pure Ti, while all atoms were allowed to fully relax until an energy convergence of 10$^{-4}$ eV and a force convergence of 0.02 eV/\AA on each atom was reached.
The Methfessel-Paxton (order 2) method was used to treat partial occupancies during relaxation runs, while the tetrahedron method with Blochl corrections was used during static runs. 
Calculations with Cr and Fe were performed spin polarized. 
Bader charges were calculated with the scheme of Henkelman et al \cite{sanville2007improved,tang2009grid,henkelman2006fast}.
The charge densities were visualized using the VESTA software \cite{momma2011vesta}.

\section{Results}\label{sec:results}

\begin{figure}
    \centering
    \includegraphics[width=1.0\linewidth]{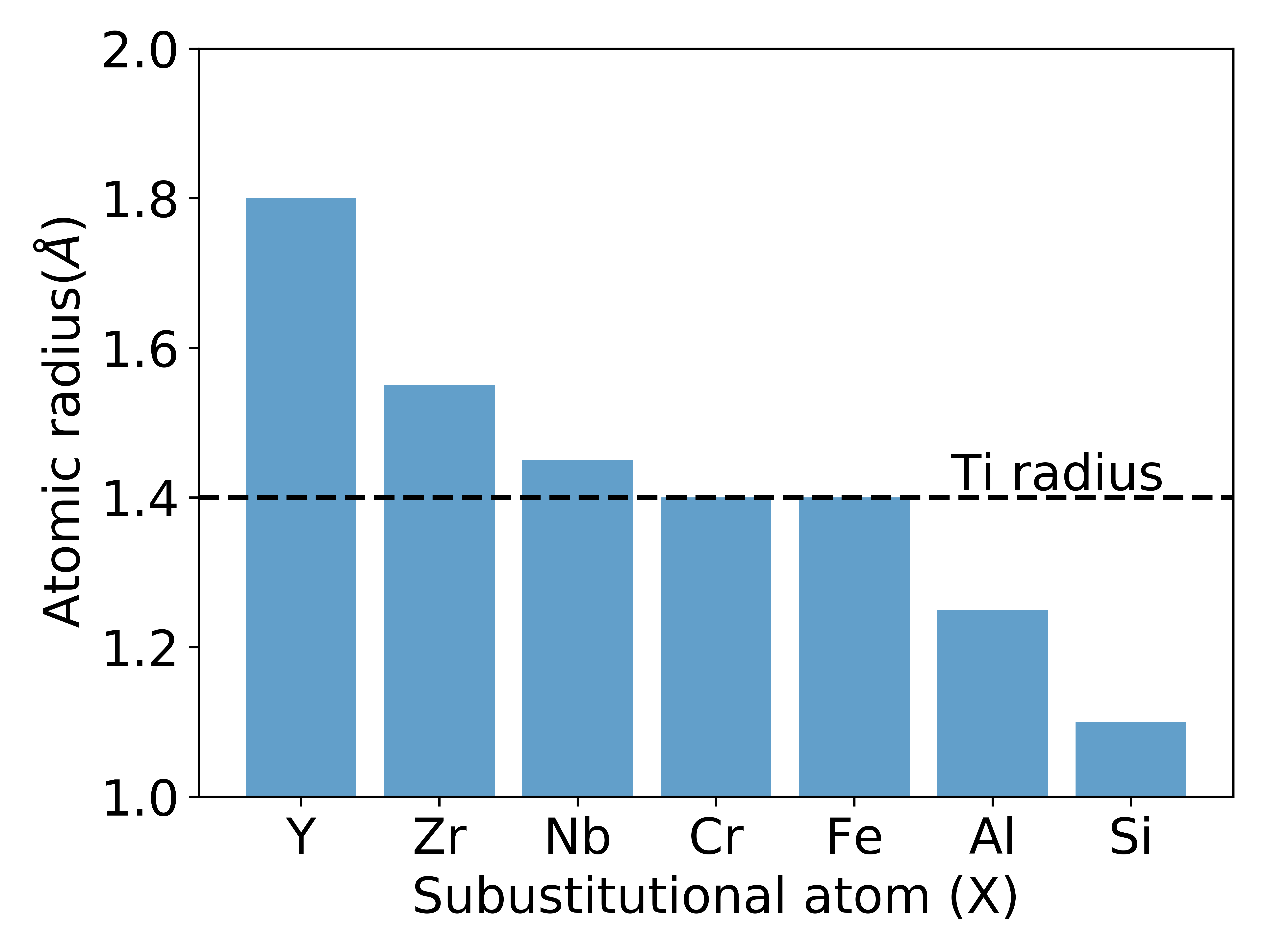}
    \caption{Atomic radii of alloying elements investigated in this study\cite{Slater1964}.}
    \label{fig:atomic_radius}
\end{figure}

We considered seven alloying elements from different parts of the periodic table with a range of atomic radii (Figure \ref{fig:atomic_radius}) and valence electronic structure to investigate trends in the interactions between dilute substitutional solutes and octahedral oxygen in hcp Ti. 
Yttrium was chosen due to its strong affinity for oxygen and its tendency to form among the most stable oxides \cite{}. 
Zirconium, Niobium and Chromium were chosen as these are refractory transition metals and are commonly combined with Ti to form high entropy alloys \cite{senkov2011microstructure,butler2017high,miracle2017critical,senkov2018development,natarajan2020crystallography}.
Crystalline Y, Zr and Nb are also able to dissolve large concentrations of interstitial oxygen \cite{abriata1986zr,paul2011first,puchala2013thermodynamics,perez2007thermodynamic}. 
Iron is a representative of a later transition metal, while Al and Si are important alloying additions that are commonly added to metals to promote the formation of protective Al$_2$O$_3$ and SiO$_2$ scales \cite{stott1995influence,chou2017influence,chou2018early}.
As is evident in Figure \ref{fig:atomic_radius}, the atomic radii of the selected elements span a range of values relative to that of Ti, decreasing upon moving to the right in the periodic table. 

\begin{figure}
    \centering
    \includegraphics[width=1.0\linewidth]{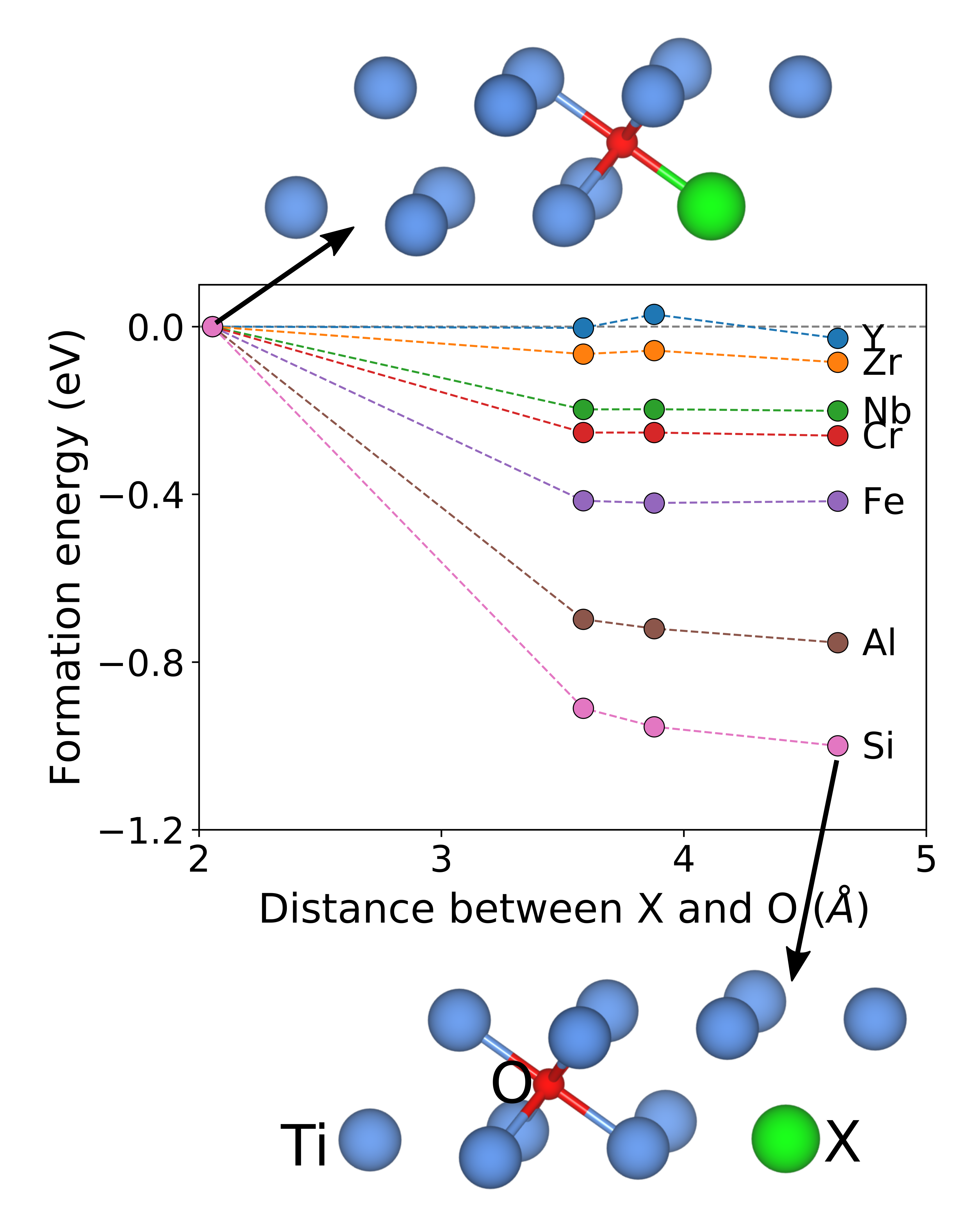}
    \caption{The variation of the energy of hcp Ti as a substitutional solute X is moved away from the nearest neighbor position of an interstitial oxygen. The reference state of the energy scale was set to the energy of the crystal when X and O are nearest neighbors at approximately 2 \AA. A $3\times3\times3$ super cell of hcp was used to calculate these energies.}
    \label{fig:energy_vs_distance}
\end{figure}

\subsection{Substitutional solute-oxygen repulsion is short ranged}\label{sec:trends}

The interaction between a substitutional solute X (Y, Zr, Nb, Cr, Fe, Al and Si) and an interstitial O in hcp Ti was determined by calculating the energy of a 3$\times$3$\times$3 super cell of the hcp unit cell as a function of the X-O separation. 
The interstitial oxygen is coordinated by six substitutional sites.
Figure \ref{fig:energy_vs_distance} shows the variation of the energy of the cell with increasing X-O distance. 
The energy for the nearest neighbor configuration in which the solute occupies a site that directly coordinates the interstitial oxygen atom at a distance of approximately 2 {\AA} is used as the reference state and is set equal to zero.
Figure \ref{fig:energy_vs_distance} shows that the energy of the super cell decreases as the solute X is moved beyond the first nearest neighbor shell of the interstitial oxygen. 
This indicates that the interaction between the solute X and oxygen is repulsive.
For most solutes, the energy varies negligibly beyond the second-nearest neighbor distance at approximately 3.5 {\AA}, indicating that the repulsion is very local and for the most part only felt within the first-nearest neighbor shell.

\begin{figure}
    \centering
    \includegraphics[width=1.0\linewidth]{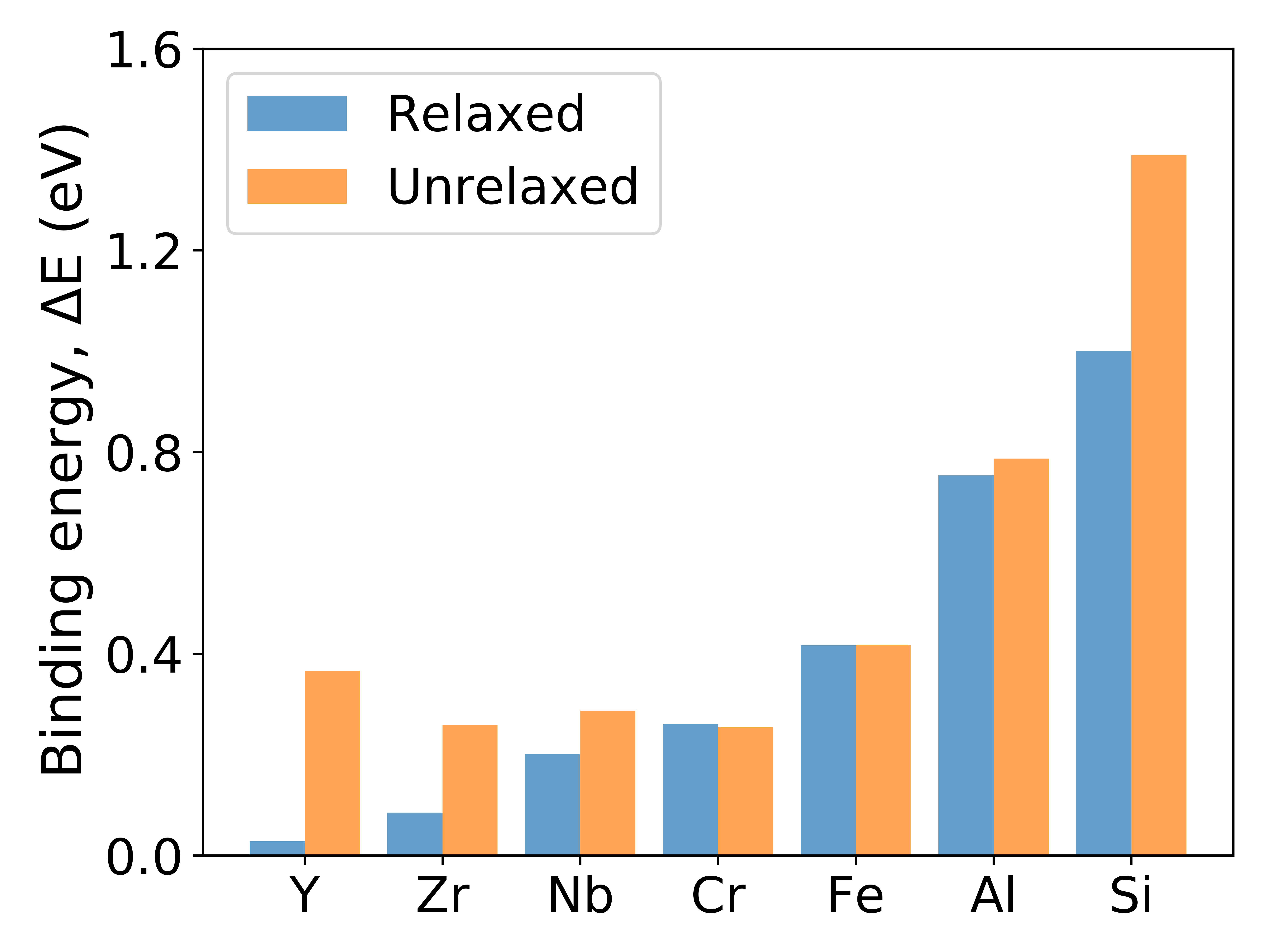}
    \caption{Binding energies, $\Delta E$, between different substitutional solutes (X = Y, Zr, Nb, Cr, Fe, Al and Si) and interstitial O in hcp Ti. The energies were calculated as the difference in the energy of a Ti super cell when X and O are nearest neighbors and the energy when X and O are fourth nearest neighbors. Positive values indicate a repulsive interaction between X and O. The relaxed $\Delta E$ values (blue) were calculated allowing for atomic relaxations at constant volume.}
    \label{fig:energy_bar_chart}
\end{figure}

The energies of Figure \ref{fig:energy_vs_distance} were calculated allowing for atomic relaxations. 
They therefore combine contributions from electronic interactions and the effects of local distortions due, for example, to a size mismatch between the solute and the Ti host. 
It is also instructive to inspect binding energies in the absence of atomic relaxations. 
Figure \ref{fig:energy_bar_chart} shows calculated binding energies $\Delta E$ for each solute, defined as the difference in energy between the nearest neighbor X-O configuration minus that of the fourth nearest neighbor configuration. 
Both relaxed (blue) and unrelaxed (orange) binding energies are shown. 
A positive binding energy signifies a repulsive interaction. 
Large differences between the relaxed and unrelaxed values are an indication of the importance of atomic relaxations in determining the binding energy $\Delta E$. 

\begin{figure}
    \centering
    \includegraphics[width=1.0\linewidth]{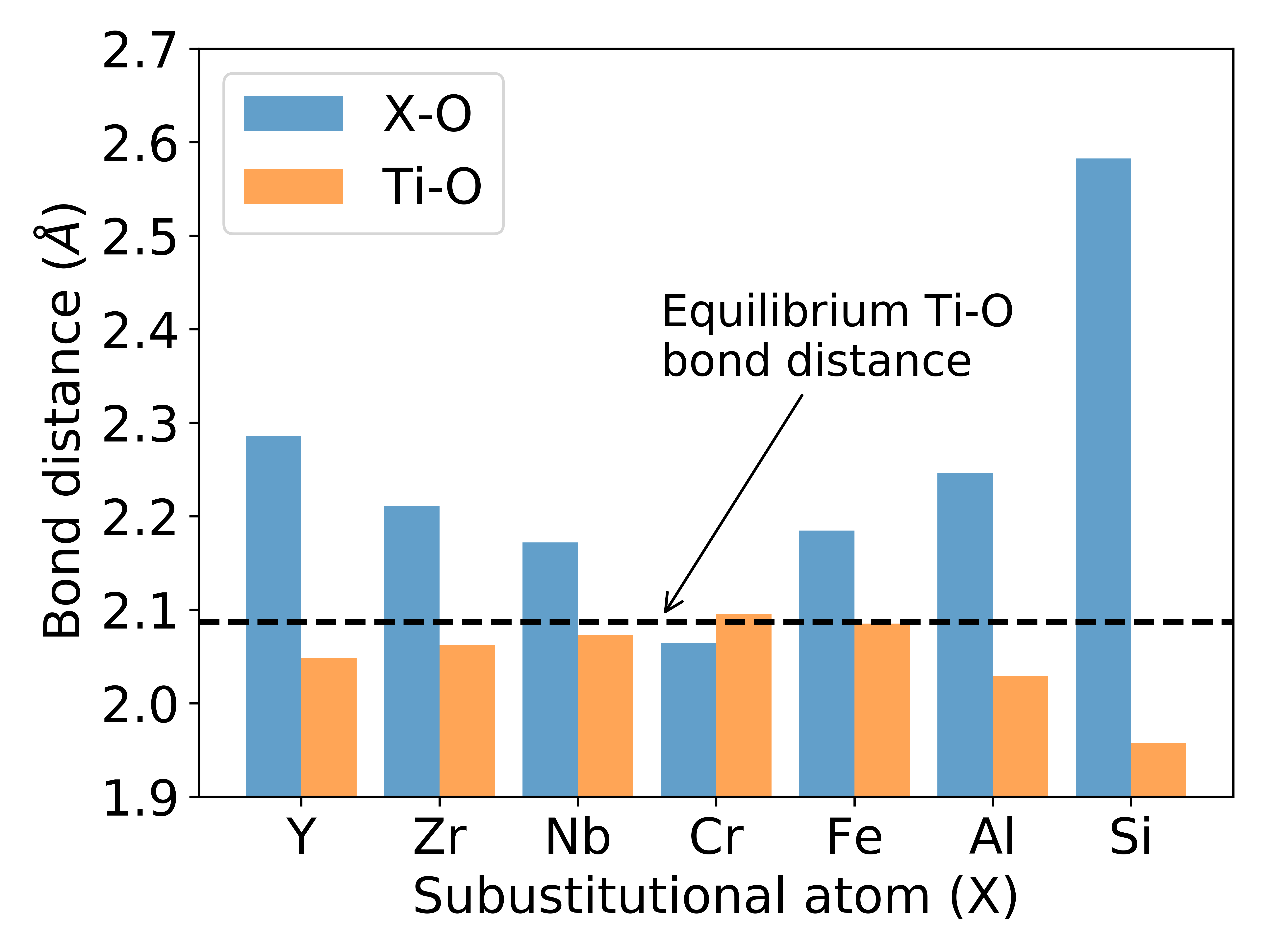}
    \caption{Relaxed bond lengths between X and O and between Ti and O when X is in the nearest neighbor shell of interstitial O.}
    \label{fig:bond_distance}
\end{figure}

Figure \ref{fig:energy_bar_chart} shows that the effects of relaxations are the most pronounced for Y, Zr and Si. 
The atomic radii of both Y and Zr are larger than the atomic radius of Ti, while that of Si is substantially smaller, as shown in Figure \ref{fig:atomic_radius}. 
In the case of Y, relaxations appear to completely undo an otherwise repulsive interaction that exists when Y and O are at their ideal positions within hcp Ti.
Figure \ref{fig:energy_bar_chart} also shows that the binding energy between oxygen and alloying elements such as Cr, Fe and Al are negligibly affected by relaxations. 
Both Cr and Fe have atomic radii that are very similar to that of Ti, while Al is slightly smaller. 

One measure of the degree of relaxation is the equilibrium X-O bond distance when X and O are nearest neighbors. 
Figure \ref{fig:bond_distance} shows the relaxed X-O bond distances and compares them to the lengths of the Ti-O bond on the opposite side of O. 
With the exception of Cr, all other alloying elements have an X-O bond length that is larger than the opposing Ti-O bond length. 
The large X-O bond lengths indicate that O is pushed away from the solute X towards the Ti on the opposite side of the octahedral interstitial site containing O, resulting in a shorted Ti-O bond. 
The relaxations are especially pronounced for Y, Zr, Al and Si. 
While Y and Zr are larger than Ti, Al and Si are smaller. 
The large Al-O and Si-O bond lengths are therefore surprising and must arise from factors other than the intrinsic size of Al and Si. 
In the next section we will show that a rehybridization of atomic-like orbitals occurs when O becomes a nearest neighbor of either a Si or Al solute.  

\subsection{An analysis of the electronic structure} \label{sec:hybridization}

\begin{figure}
    \centering
    \includegraphics[width=1.0\linewidth]{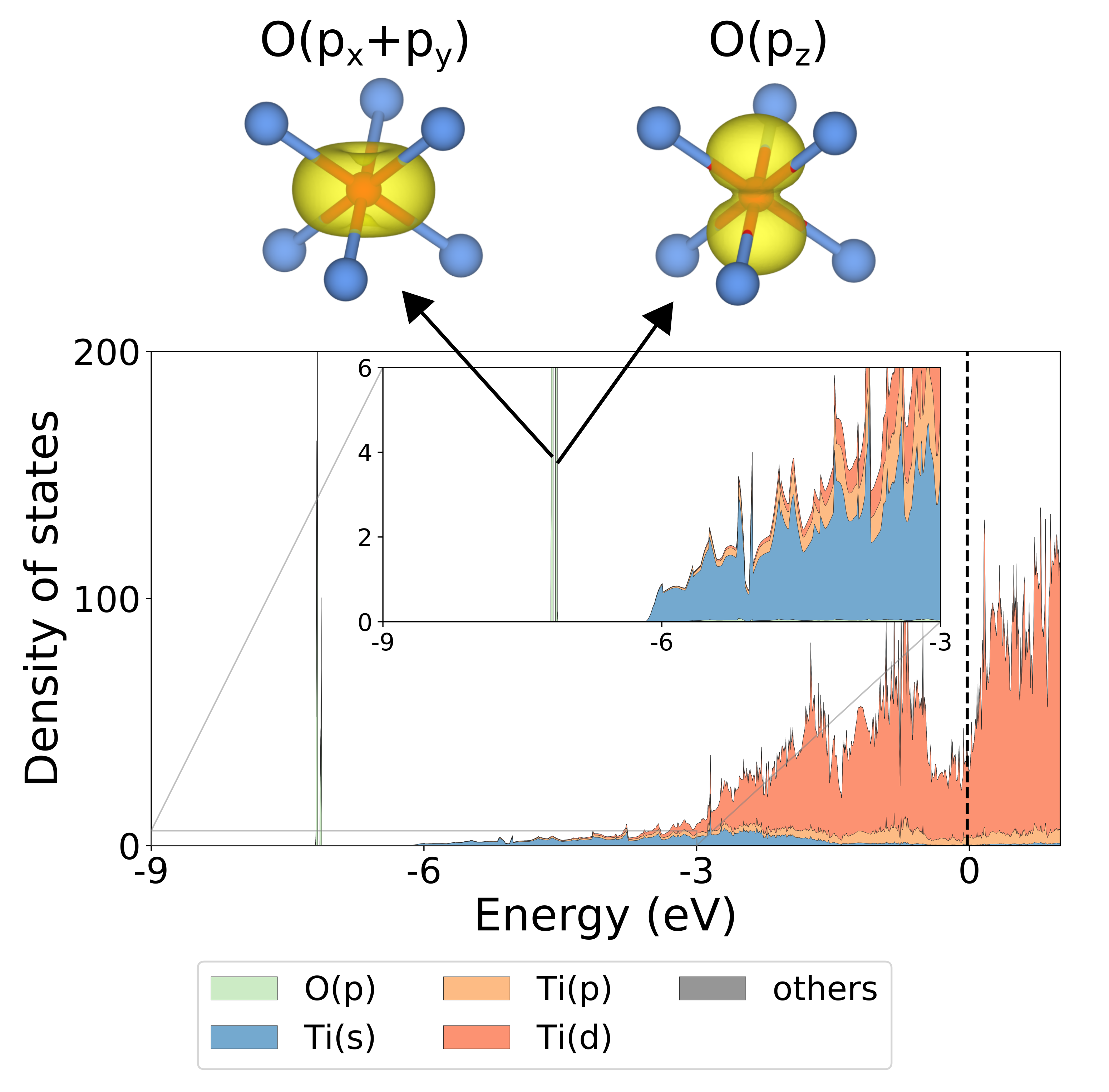}
    \caption{The electronic densities of states (DOS) of a super cell of hcp Ti containing a single interstitial oxygen. The insets show the charge densities corresponding to the narrow spikes in the DOS at low energy.}
    \label{fig:Ti_O_DOS}
\end{figure}

We next analyze changes in electronic structure as an alloying element X is brought into the nearest neighbor shell of an oxygen atom residing in an octahedrally coordinated interstitial site of hcp Ti.
It is instructive to first consider the electronic density of states (DOS) of a super cell of hcp Ti containing one interstitial oxygen in the absence of any other solutes.
This is shown in Figure \ref{fig:Ti_O_DOS}, where the position of the Fermi level is denoted by the vertical dashed line.
The electronic states having Ti $s$ and $d$ character are shown in blue and orange, respectively. 
The valence $p$-levels of oxygen have intrinsic energies that are far removed from the valence states of Ti derived from the 4$s$ and 3$d$ levels. 
Hence, there is very little hybridization between the oxygen $p$ levels and the valence electrons of Ti, resulting in highly localized states around oxygen within a very narrow energy range as shown in the calculated density of states plot of Figure \ref{fig:Ti_O_DOS}.
Two very closely spaced peaks derived primarily from oxygen $p$ states are evident. 
One resembles a $p_{z}$-like state, which is aligned along the $c$ axis of the hcp crystal. 
The other resembles a superposition of the $p_{x}$ and $p_{y}$ orbitals that form a doughnut shaped charge density parallel to the basal plane of the hcp crystal. 
Their charge densities are shown as insets in Figure \ref{fig:Ti_O_DOS}.

\begin{figure}
    \centering
    \includegraphics[width=1.0\linewidth]{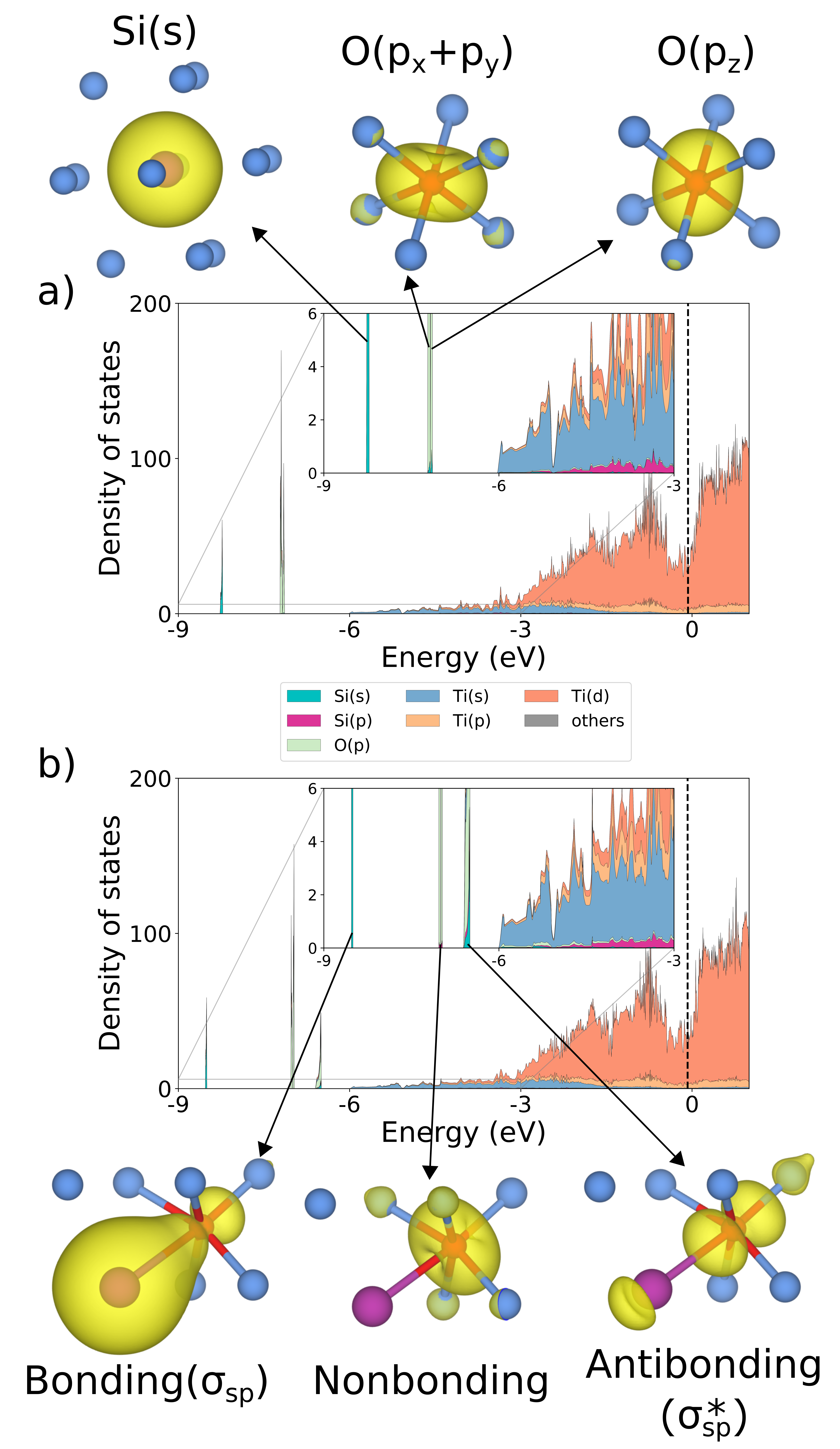}
    \caption{(a) The DOS of a Ti super cell containing and single substitutional Si in the fourth nearest neighbor shell of an interstitial O. (b) The DOS of the same crystal when Si is in the nearest neighbor shell of an interstitial O. }
    \label{fig:Si-O_DOS}
\end{figure}

Figure \ref{fig:Si-O_DOS}a shows the DOS for a super cell of Ti containing an interstitial oxygen and a substitutional Si placed in the fourth nearest neighbor shell of the oxygen. 
While the Si $p$ states hybridize with the Ti valence states, the Si 2$s$ states do not and remain highly localized as illustrated in  Figure \ref{fig:Si-O_DOS}a. 
Also evident in Figure \ref{fig:Si-O_DOS}a are the oxygen $p$ states, which are very similar to those of oxygen in pure Ti without a Si impurity. 
%with a doughnut shaped charge density and a $p_z$-like orbital aligned along the $c$-axis.
The energy of the Si $s$-state is lower than that of the oxygen $p$ states. 

When Si is moved to the nearest neighbor shell of oxygen, the local $s$-like state of Si interacts strongly with the $p$-like orbitals of oxygen.
This is evident in the calculated DOS of Figure \ref{fig:Si-O_DOS}b for a super cell of Ti in which a Si solute is a nearest neighbor of oxygen.
Three separated peaks emerge, each with very distinct charge densities, as shown in Figure \ref{fig:Si-O_DOS}b.
The lowest peak at approximately -8.5 eV resembles a bonding like orbital that arises from the hybridization between the Si $s$ state and an O $p$ state, with its axis aligned along the Si-O bond. 
The next peak at approximately -7 eV has a doughnut shaped charge density arising from a superposition of $p$-like orbitals. 
In contrast to an isolated oxygen atom (Figure \ref{fig:Ti_O_DOS} and Figure \ref{fig:Si-O_DOS}a), however, the doughnut shaped charge density has rotated to become perpendicular to the Si-O bond. 
This state appears to be a non-bonding orbital as it does not contain any discernable states residing on the Si atom. 
The highest peak at approximately -6.5 eV resembles the anti bonding state of the hybridization between the Si $s$ state and an O $p$ state.
This state has charge density on both the Si and O, but the charge density is pushed away from the center of the Si-O bond. 

\begin{figure}
    \centering
    \includegraphics[width=1.0\linewidth]{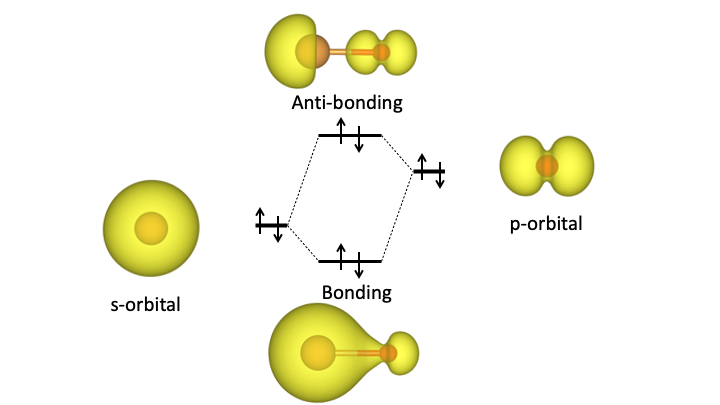}
    \caption{Schematic hybridization diagram and charge densities that emerge when an $s$ orbital and a $p$ orbital combine to form bonding and anti-bonding orbitals.}
    \label{fig:hybridization}
\end{figure}

For comparison, Figure \ref{fig:hybridization} shows a schematic hybridization diagram between an $s$ orbital (on the atom on the left) and a $p$ orbital (on the atom on the right). 
In the diagram of Figure \ref{fig:hybridization}, the intrinsic energy of the $s$ level is lower than that of the $p$ level, similar to the $s$-level of Si and the $p$ levels of O when the two atoms are separated from each other in a Ti crystal. 
In this scenario, the bonding state is dominated by the $s$ level with excess charge accumulating along the bond. 
The antibonding state has more of the higher energy $p$ character with a depletion of charge between the two atoms.
The charge densities of Figure \ref{fig:hybridization} are very similar to those of the states labeled as bonding and anti-bonding in Figure \ref{fig:Si-O_DOS}b.

\begin{figure}
    \centering
    \includegraphics[width=1.0\linewidth]{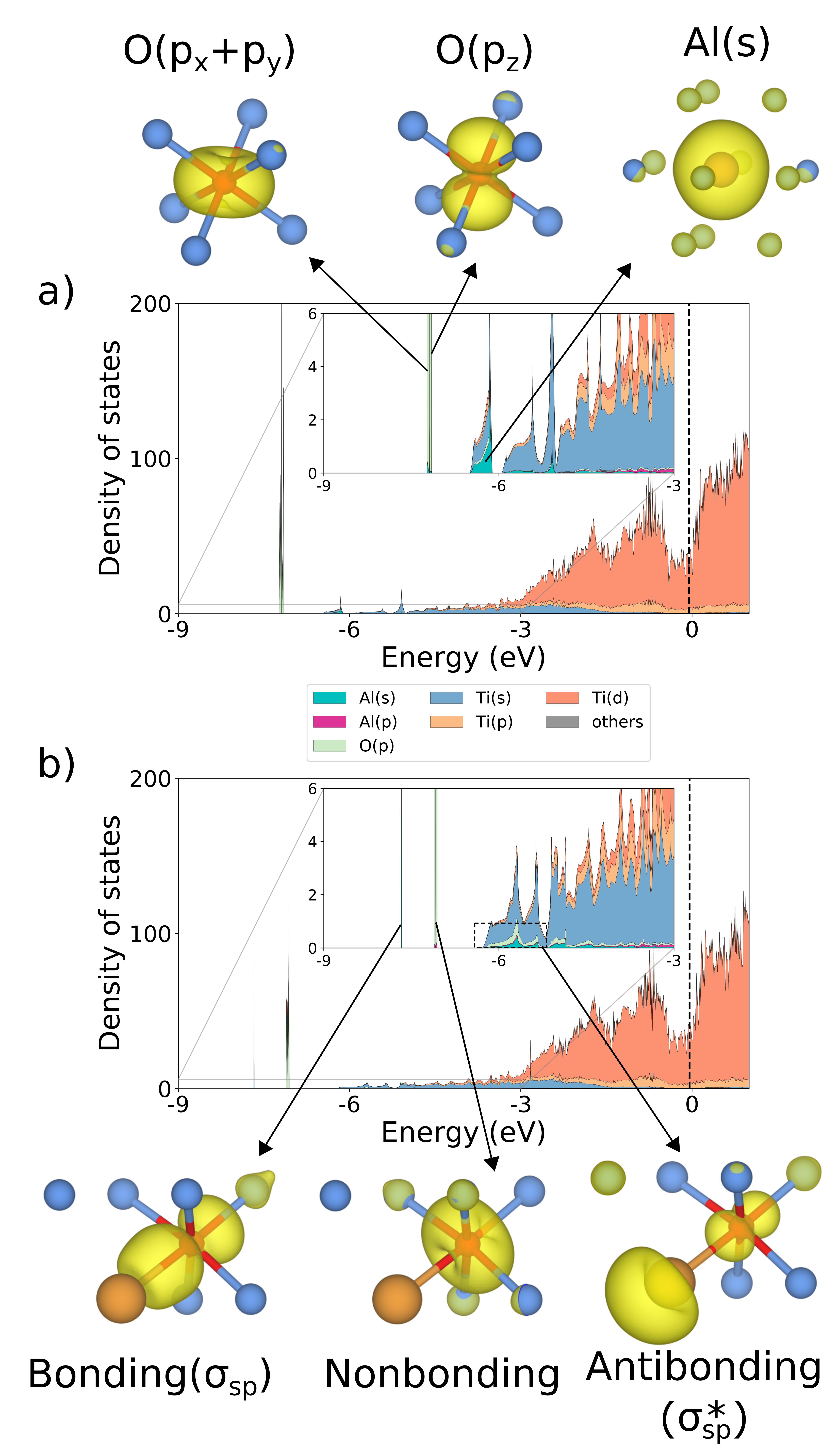}
    \caption{(a) The DOS of a Ti super cell containing a single substitutional Al in the fourth nearest neighbor shell of an interstitial O. (b) The DOS of the same crystal when Al is in the nearest neighbor shell of an interstitial O.}
    \label{fig:Al-O_DOS}
\end{figure}

Another scenario is also possible in which the intrinsic energy of the $p$ level is lower than that of the $s$ level.  
In this case the bonding state has more $p$ character and the antibonding state is more dominated by the $s$ state. 
This scenario occurs when Al and O are nearest neighbors in a Ti host as was revealed in Ref \cite{gunda2020understanding} and shown in Figure \ref{fig:Al-O_DOS}. 

Figure \ref{fig:Al-O_DOS}a shows the DOS of a super cell of Ti with a substitutional Al and interstitial oxygen beyond their nearest neighbor shells. 
In contrast to Si, the Al $s$ levels are above those of the oxygen $p$ levels and are somewhat mixed in with the Ti $s$ levels.
When Al is brought into the nearest neighbor shell of the oxygen, a hybridization between the Al $s$ and O $p$ levels becomes evident (Figure \ref{fig:Al-O_DOS}b). 
A bonding state emerges around -7.7 eV. 
The charge density associated with this state now has more of a $p$ character since the intrinsic energy of the O $p$ level is below that of the Al $s$ level with which it hybridizes. The next levels around -7.1 eV again have the appearance of non-bonding states, with a doughnut shaped charge density that is oriented perpendicular to the Al-O bond. 
The anti-bonding states are less localized as they also mix with the Ti $s$ levels. 
Their charge density in the vicinity of the Al-O bond of these states shows features of an $s$ orbital on Al and a $p$ orbital on O, but with a clear depletion of charge at the center of the bond. 

\begin{figure}
    \centering
    \includegraphics[width=1.0\linewidth]{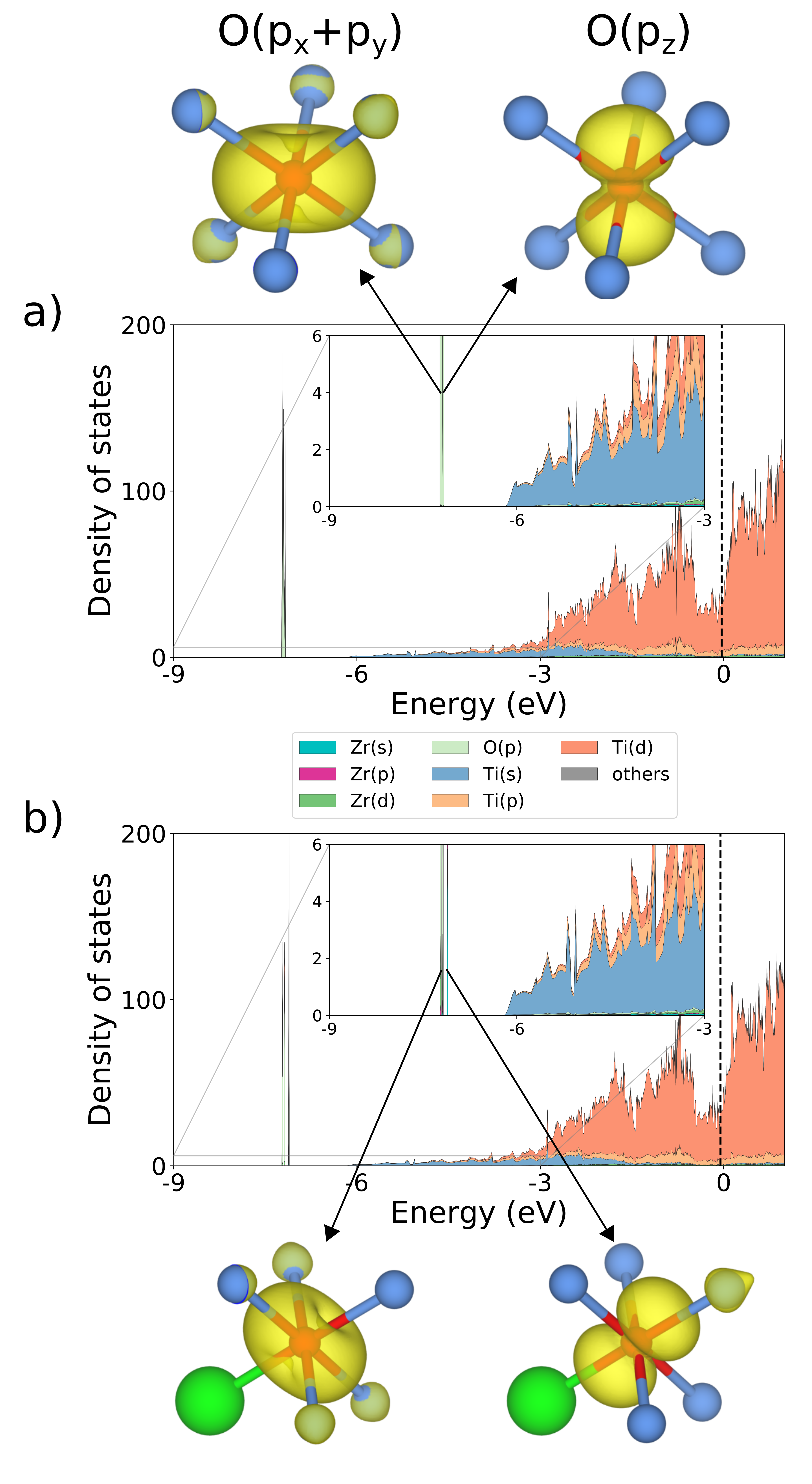}
    \caption{(a) The DOS of a Ti super cell containing a single substitutional Zr in the fourth nearest neighbor shell of an interstitial O. (b) The DOS of the same crystal when Zr is in the nearest neighbor shell of an interstitial O.}
    \label{fig:Zr-O_DOS}
\end{figure}

\begin{figure}
    \centering
    \includegraphics[width=1.0\linewidth]{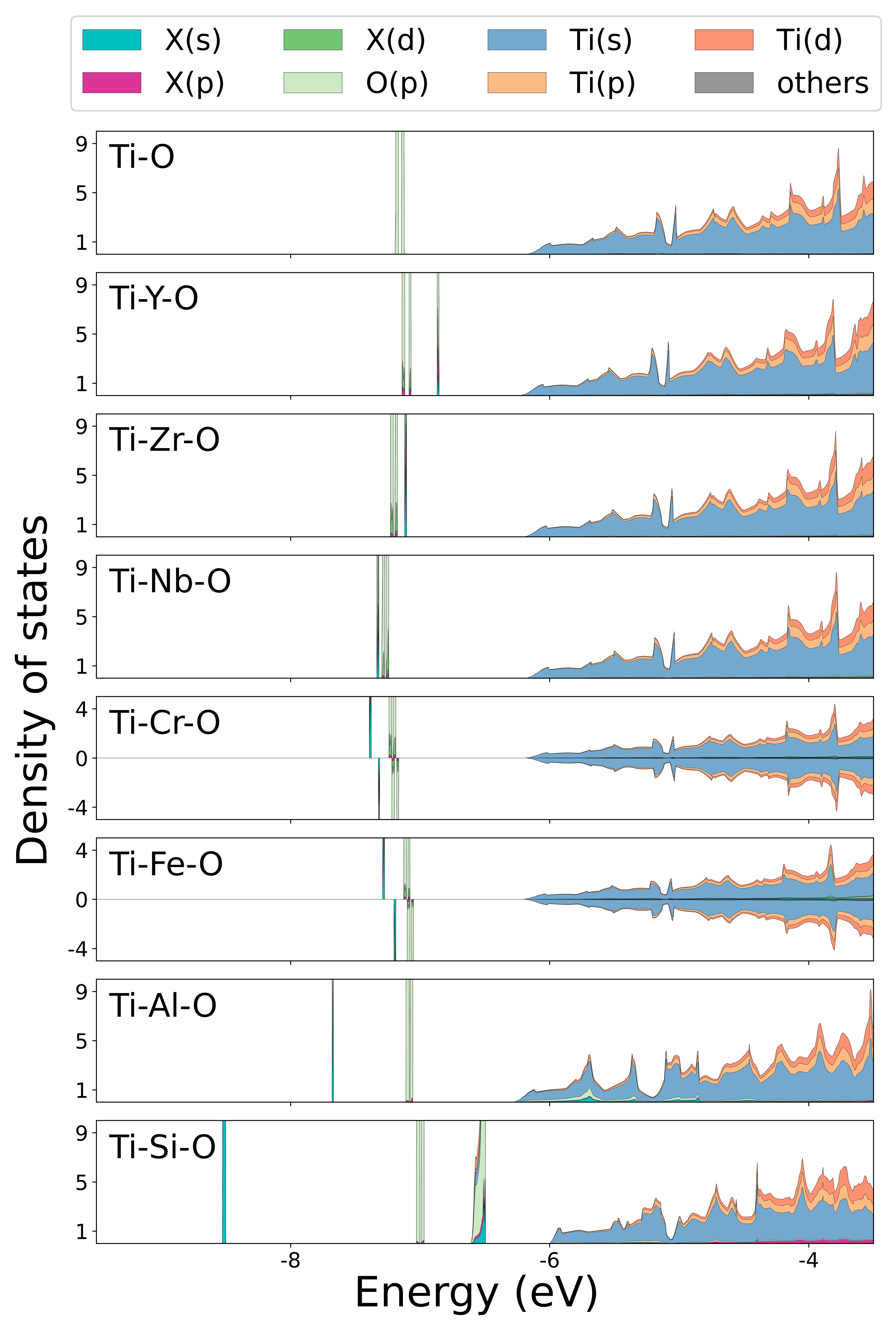}
    \caption{A compilation of the DOS of a Ti super cell containing a nearest neighbor X-O pair (X=Y, Zr, Nb, Cr, Fe, Al and Si). The DOS of a Ti super cell with a single oxygen and no solutes is also shown (top) for comparison. Both the spin up and spin down DOS are shown for Ti super cells containing Cr and Fe as there is a slight degree of spin polarization. }
    \label{fig:Ti_X_O_DOS}
\end{figure}

For the transition metal alloying elements (i.e. Y, Zr, Nb, Cr and Fe), the effects of hybridization are less pronounced. 
We focus on the Ti-Zr-O system (Figure \ref{fig:Zr-O_DOS}) as a representative example. 
The DOS plots of the other transition metal solutes are similar (Figure \ref{fig:Ti_X_O_DOS} and supporting information). 
When Zr is in the fourth nearest neighbor shell of O, the oxygen $p$ levels again resemble those of an isolated oxygen atom in pure Ti, as is evident in Figure \ref{fig:Zr-O_DOS}a. 
In contrast to Si and Al, the Zr valence states are very similar to those of Ti. 
The Zr valence $s$ and $d$ states are therefore mixed in with those of Ti.
The Zr $s$ levels, for example, have energies that are above those of the oxygen $p$ levels and the difference that separates them is significantly larger than in the case of Si and Al. 
Hence, the degree of hybridization will be less pronounced when Zr moves into the nearest neighbor shell of oxygen as can be seen in Figure \ref{fig:Zr-O_DOS}b.
The oxygen $p$ levels of an isolated oxygen (Figure \ref{fig:Zr-O_DOS}a) are only slightly modified when Zr becomes a nearest neighbor (Figure \ref{fig:Zr-O_DOS}b).
Two nearly degenerate peaks are evident whose combined charge density has a doughnut shape that is almost perpendicular to the X-O bond. 
A third peak at a slightly higher energy has a charge density of a $p$-like orbital that is aligned approximately parallel to the X-O bond. 

Figure \ref{fig:Ti_X_O_DOS} collects the DOS of all the X-O couples when the solute X is within the nearest neighbor shell of oxygen.
The zero on the energy axis of each DOS plot was chosen such that the deep Ti 3$s$ level aligns with that of pure Ti. 
This makes it possible to compare the extent to which different solutes hybridize with the localized O $p$ levels.
The effects of transition metal solutes on the oxygen $p$ levels are substantially smaller than those of Al and Si. 
Figure \ref{fig:Ti_X_O_DOS} shows that Nb has the smallest effect on the oxygen levels. 
For Y and Zr, the energy of the peak corresponding to a $p$-like orbital aligned parallel to the X-O bond is higher than the nearly degenerate pair of peaks with a doughnut shaped charge density, while for Cr and Fe, these peaks are reversed. 
Whether the $p$-like orbital parallel to the X-O bond has a higher or lower energy than the degenerate $p$-states with a doughnut charge density appears to be correlated with the atomic radius of the solute relative to that of Ti (Figure \ref{fig:atomic_radius}): for the larger solutes (Y, Zr) the energy is higher while for the smaller solutes (Nb, Cr and Fe) the energy is lower. 

\subsection{Bader charges of O and X}\label{sec:bader}

The energies of the $p$ levels of dissolved oxygen in hcp Ti reside well below the Fermi level and the electronic valence states derived from Ti 4$s$ and 3$d$ levels. 
Hence, the O $p$ states will draw electrons from the surrounding Ti host, leading to a filled shell configuration on oxygen and a local accumulation of charge. 
An accumulation of charge on oxygen is confirmed with calculated Bader charges, which predict a value of -1.4 for an oxygen atom in a $3\times3\times3$ super cell of hcp Ti. 
This suggests that the Ti host acts as an electron reservoir from which the dissolved oxygen can draw electrons to achieve a closed-shell configuration.
It should be noted that the charge of an atom in a solid is not a well-defined quantity. 
Bader charges are one way of qualitatively assessing whether an atom accumulates charge or donates it. 
While the calculated Bader charge of oxygen in Ti is -1.4, which contrasts with a charge of -2 for a closed shell, the calculated DOS plots clearly indicated a closed shell configuration on the interstitial oxygen with all three valence $p$ states residing well below the Fermi level and therefore fully occupied. 

\begin{figure}
    \centering
    \includegraphics[width=1.0\linewidth]{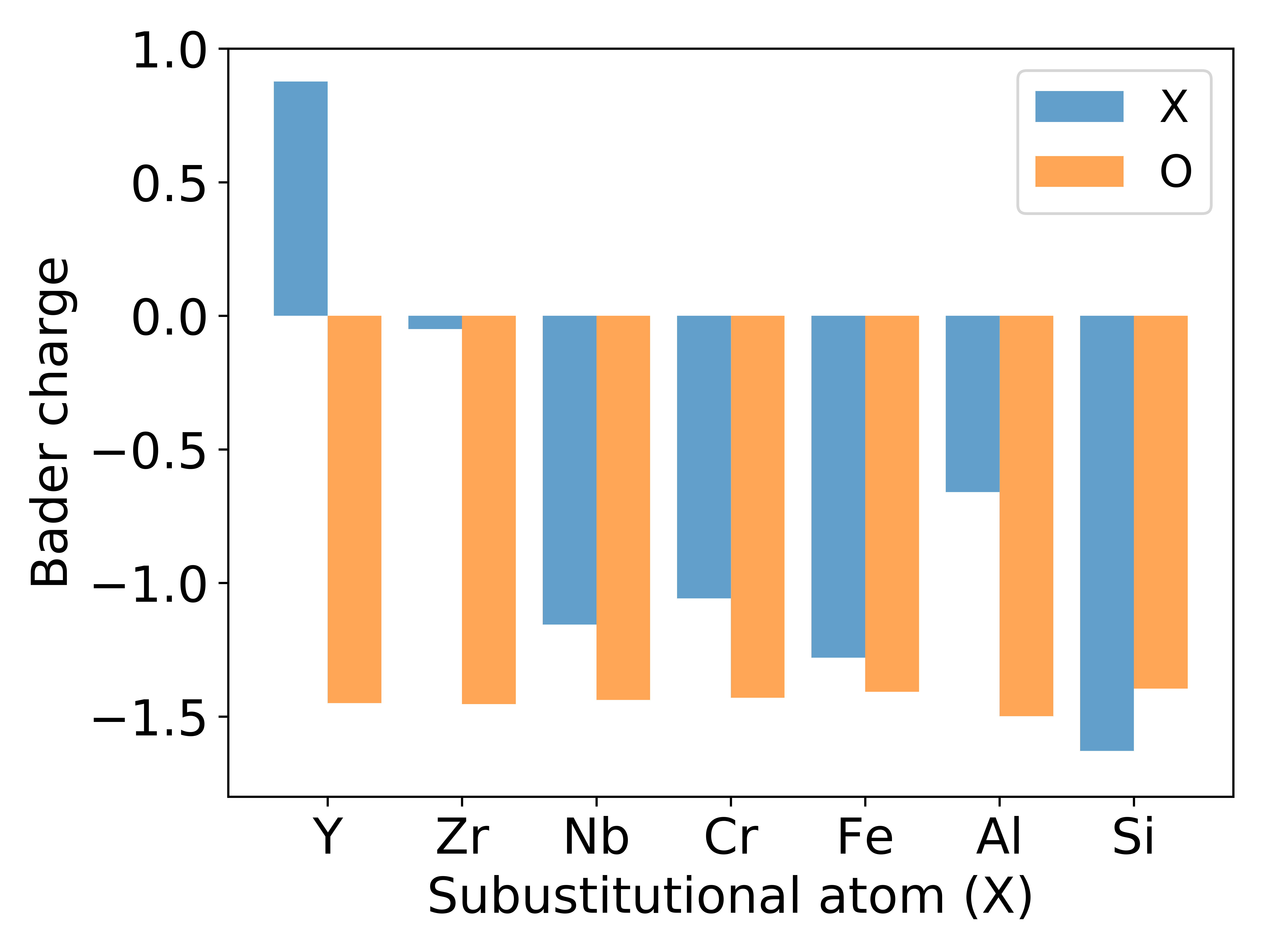}
    \caption{Calculated Bader charges on different solutes and O in a super cell of hcp Ti when the solute and oxygen are nearest neighbors.}
    \label{fig:bader_charge}
\end{figure}

The alloying elements X may also donate or accept charge from the Ti host depending on the intrinsic energies of their valence states relative to those of Ti. 
Figure \ref{fig:bader_charge} collects the calculated Bader charges for the different alloying elements when they are in the nearest neighbor shell of an interstitial oxygen. 
The Bader charges of the solutes do not change significantly when they are in the fourth nearest neighbor position from oxygen and thereby largely surrounded by Ti. 
With the exception of Y and Zr, all other alloying elements have a negative Bader charge. 
Y solutes have a net positive charge, suggesting that they donate charge to the Ti host. 
Zr solutes have a negligible Bader charge, which is consistent with the fact that Ti and Zr are both group 4 elements and have a similar valence electron structures.

The predicted Bader charges suggest the importance of electrostatic interactions between O and X. 
However, since the Ti host is metallic, any Coulomb interactions between X and O will be strongly screened by the free electrons of the host. 
It is only when X and O are nearest neighbors, with no intervening Ti, that the net charges on both X and O may lead to strong Coulomb interactions.

\section{Discussion}\label{sec:discussion}

Alloying elements are often added to Ti to reduce its oxygen solubility. 
A first-principles study by Wu and Trinkle \cite{wu2013solute} showed that almost any element from the periodic table that is substitutionally dissolved in hcp Ti will repel interstitial oxygen. 
Our calculations on a subset of substitutional alloying elements confirm these predictions. 

While a simple explanation for a repulsive or attractive interaction between solutes in metals is often difficult to discern from the results of first-principles electronic structure calculations, there are two distinct types of interactions between substitutional solutes and interstitial oxygen in hcp Ti that can be identified that partially explain the trends in the calculated binding energies $\Delta E$ of Figure \ref{fig:energy_bar_chart}.
The first is a short-ranged {\it closed-shell} repulsion that emerges from the hybridization between oxygen $p$ and solute valence states. 
The second is {\it electrostatic} in nature and arises from a redistribution of electron density between the Ti host and the solutes that make both the substitutional solute and interstitial oxygen charged. 
We elaborate further on these two interactions. 

A remarkable property of Ti is that it does not appear to interact strongly with isolated interstitial oxygen. 
The energy of the oxygen $p$ levels are far removed from those of the Ti 4$s$ and 3$d$ levels. 
The DOS peaks associated with the oxygen $p$ levels (Figure \ref{fig:Ti_O_DOS}) are very sharp, which is indicative of localized, atomic-like orbitals that hybridize negligibly with the surrounding Ti host.
As described in Section \ref{sec:hybridization}, the placement of a substitutional solute, X, in the first-nearest neighbor shell of an interstitial oxygen atom alters the positions of the oxygen $p$ peaks and the shapes of their associated charge densities. 

Figure \ref{fig:Ti_X_O_DOS} shows that Al and Si induce the largest degree of hybridization.
The calculated DOS and charge densities of Figures \ref{fig:Si-O_DOS} and \ref{fig:Al-O_DOS}
show that the hybridization is primarily between the oxygen $p$ orbitals and the $s$ states of Al or Si.
Furthermore, the hybridization between O and Al/Si within Ti leads to clearly discernable bonding and anti-bonding states. 
For transition metal solutes, such as Y, Zr, Nb, Cr and Fe, the effect is not especially pronounced. 
While similar hybridization phenomena may also occur between oxygen and transition metal solutes, it is less evident from an inspection of DOS plots in Figure \ref{fig:Ti_X_O_DOS}. 
Notable though is that the Nb-O nearest neighbor pair in hcp Ti leads to the least degree of splitting of the original O $p$ levels. 
Among the transition metals considered here, Y produces the largest splitting of the O $p$ levels.

The localized bonding and anti-bonding states on the Si-O and Al-O pairs are well below the Fermi level of the Ti host and are consequently fully occupied. 
The deep lying $s$ and $p$ states should, therefore, behave similarly to those of closed-shell atoms as they approach each other. 
Since both the bonding and anti-bonding states are filled, there is no energetic benefit to hybridize and form an attractive bond. 
Furthermore, closed-shell atoms repel each other at short distances due to the Pauli exclusion principle as their closed-shell electron clouds begin to overlap.

The existence of a closed-shell repulsion between dissolved oxygen and solutes such as Al and Si can be inferred from the relaxed X-O bond lengths. 
The bond lengths collected in Figure \ref{fig:bond_distance} show that, while Al and Si are smaller than Ti (Figure \ref{fig:atomic_radius}), the relaxed Al-O and Si-O distances are significantly larger than a nearest neighbor Ti-O bond. 
We interpret this as arising from a strong repulsion between the closed-shell O $p$ and Al/Si $s$ orbitals. 
Due to the constraint of the crystal, any increase in the X-O bond length needs to be accommodated by a shortening of the surrounding Ti-O bond distances. 
The Ti-O bond lengths opposite to those of the Al-O or Si-O bonds are very short (Figure \ref{fig:bond_distance}), especially when compared to the Ti-O bond distances when O is next to a transition metal solute such as Nb, Cr or Fe.
The degree with which these Ti-O bonds are compressed is a measure of the strength of the closed-shell O-X repulsion since the X-O bond length will continue to relax (i.e. lengthen) until it is countered by an equal force by the opposing Ti-O bond. 
Based on these arguments, we concluded that the closed-shell repulsion is especially strong for Al-O and Si-O bonds. 

Similar arguments may hold for Y and Zr, however, both Y and Zr are larger than Ti and it may simply be their larger size that is responsible for the increased Y-O and Zr-O distances (Figure \ref{fig:bond_distance}).
It should be noted that in the case of Y, there is a sizable splitting of the oxygen $p$ orbitals when Y and O become nearest neighbors, as is evident in Figure \ref{fig:Ti_X_O_DOS}. 
Furthermore, the energy of the oxygen $p$ orbital pointing to Y or Zr is raised relative to the $p$ orbitals that form a doughnut shaped charge density perpendicular to the X-O bond (See Figure \ref{fig:Zr-O_DOS}b for Zr). 
While the nature of any hybridization between O and Y or Zr is difficult to discern, the increase in the energy of the $p$ orbital pointing towards Y or Zr may be due its overlap with charge density on the larger Y and Zr atoms.
We find, for example, that the energy of this localized O $p$-like state is very sensitive to the O-Y distance and decreases with increasing bond distance, again suggestive of a closed-shell like repulsion at short distances. 
This is evident in the DOS plots of the unrelaxed structures shown in Supporting Information. 

Inspection of Figure \ref{fig:bond_distance} suggests that the repulsion between Cr and O due to hybridization effects is negligible since the Cr-O bond length is very similar to the opposing Ti-O bond length. 
Furthermore, the unrelaxed and relaxed binding energies for Cr-O pairs are very similar (Figure \ref{fig:energy_bar_chart}), indicating that relaxations are negligible when Cr is in the nearest neighbor shell of O. 

While relaxations do not play a significant role in determining the interaction between Cr and O in hcp Ti, the binding energy $\Delta E$ is nevertheless still sizable and positive (Figure \ref{fig:energy_bar_chart}).
This suggests that size effects and/or closed-shell repulsion cannot be the only factors contributing to the X-O binding energies of Figure \ref{fig:energy_bar_chart}.
The calculated Bader charges of Figure \ref{fig:bader_charge} point to the importance of electrostatic interactions.
Figure \ref{fig:bader_charge} shows that both Cr and O have a negative charge. 
While the surrounding metallic Ti host, with its high density of itinerant electrons at the Fermi level, can screen the negative charges when Cr and O are at large distances from each other, it will be less effective when they are nearest neighbors.
As nearest neighbors, the negatively charged Cr and O can be expected to repel each other through Coulomb interactions.  

Electrostatic interactions should also contribute to the binding energies between oxygen and the other solute elements. 
Al and Si, where hybridization and close-shell repulsion is strong, will have an additional repulsion due to the accumulation of negative charges on both O and Si/Al. 
The negative Bader charge on Si is especially large, while that on Al is significantly smaller.

The Y solutes have a binding energy with O that is close to zero, in spite of the fact that they strive for a large Y-O distance that compresses the opposing Ti-O bond length and thereby incurs an energy penalty. 
Nevertheless, the Bader charge on Y is positive, while that on O is negative (Figure \ref{fig:bader_charge}).
This should lead to an attractive Coulomb interaction between the oppositely charged solutes at short distances, which apparently compensates for the energy penalty of the larger Y-O bond length. 
According to Figure \ref{fig:bader_charge}, Zr has almost no net accumulated charge, which is consistent with the fact that Zr and Ti have very similar valence electron structures. 
Short-range Coulomb interactions between Zr and O are likely not significant. 
The positive (repulsive) binding energy of a Zr-O bond therefore has its origin in large part from the increased Zr-O distance and compressed Ti-O distances. 

The binding energy between Fe and O appears to be dominated by electrostatic interactions. 
While the Fe-O bond does lengthen relative to that of the Ti-O bond length in the absence of solutes (Figure \ref{fig:bond_distance}), the effect of relaxations on the binding energy is minimal (Figure \ref{fig:energy_bar_chart}).
The Bader charge on Fe is larger than that of Cr, which is consistent with a larger binding energy $\Delta E$ between Fe and O compared to that of Cr. 
The binding energy between Nb and O is more difficult to explain. 
Its binding energy with oxygen is slightly smaller than that of Cr and almost half that of Fe. 
Nevertheless, it has a negative Bader charge between that of Cr and Fe and also undergoes some degree of relaxation resulting in a shortened opposing Ti-O bond. 
This suggests that the Nb-O binding energy should be more similar to that of Fe.
The difference between the Fe-O and Nb-O binding energies may arise from more subtle electronic effects. 
For example, as is clear in Figure \ref{fig:Ti_X_O_DOS}, Nb has the smallest effect on the O $p$ levels. 
Fe in contrast does lead to peak splitting that is qualitatively similar to that of Al, but to a much lesser extent. 

In summary, the trends in the binding energies between dissolved X and O can be explained to a large extent by two types of interactions that are rooted in the electronic structure of the solutes and the Ti host. 
One derives from a closed-shell repulsion that emerges due to the filling of bonding and anti-bonding states that form from localized atomic-like orbitals on the two solutes. 
While the closed-shell repulsion is short-ranged it appears to cause a lengthening of the X-O bond that is correlated with the degree of hybridization. 
Due to the constraints of the crystal, the lengthening of the X-O bond leads to a distortion of surrounding Ti-O bonds and therefore an energy penalty. 
A second interaction is a short-ranged Coulomb interaction due to a local accumulation/depletion of charge on the solute. 
Taken together, both interactions produce a net repulsive interaction between almost all substitutional solutes and interstitial oxygen. 
Other, more subtle interactions undoubtedly also play a role, but these are more difficult to identify. 

As a final note, we point out that the alloying elements considered in this study interact very differently with Ti at non-dilute concentrations. 
To some degree, the behavior at non-dilute concentrations can be correlated with the atomic size differences and the calculated Bader charges.
The Ti-Y phase diagram consists of large two-phase regions with very little Y solubility in hcp and bcc Ti \cite{gupta2009ni}. 
This is likely due to the large size mismatch between Ti and Y. 
Zr and Ti are completely miscible, while Nb can dissolve into hcp Ti up to several atom percent below the Ti hcp-bcc transition temperature before forming a large bcc solid solution. 
In contrast to the early transition metals, Cr, Fe, Al and Si are strong compound formers with Ti. 
Cr combines with Ti to form the stable Cr$_2$Ti Laves phase \cite{natarajan2020crystallography} while Fe has minimal solubility in hcp Ti and forms the stable TiFe intermetallic with the B2 structure\cite{murray1981fe}.
The solubility of Si in hcp Ti is small and quickly leads to the formation of Ti$_3$Si, a line compound, whereas the solubility of Al in hcp is much larger and leads to the formation of Ti$_3$Al, which exists over a range of compositions \cite{schuster2006reassessment,fiore2016assessment}.
It is likely no coincidence that this trend correlates with a steady increase in the X-O binding energy and an appreciable negative Bader charge on the solutes X upon going to the right in the periodic table. 
Nb is again an exception, with a negative Bader charge in the dilute limit, but nevertheless forming a solid solution with Ti. 

\section{Conclusion}
We have used first-principles electronic structure methods to shed light on the nature of the interactions between oxygen and dilute solutes in hcp Ti. 
By focusing on representative elements from different parts of the periodic table, we have identified two distinct interaction types that in combination produce a short-ranged repulsion between substitutional solutes in hcp Ti and interstitial oxygen. 
One interaction is similar to the repulsion between closed-shell atoms at short distances, and becomes more pronounced with an increase in the degree of hybridization between atomic-like orbitals on oxygen and the solute. 
Another is electrostatic in nature and arises from a transfer of electron density to the solutes from the Ti host. 
We expect that the interactions among solutes identified in this study may also play an important role in many other alloys, especially multi-principle element alloys made of refractory metals that tend to dissolve high concentrations of interstitial species such as O, N and C. 

\section{Acknowledgement}

N. S. Harsha Gunda acknowledges Dr. J. Vinckeviciute for helpful discussions on visualizing the charge densities shown in the present work. 
This work was supported by the National Science Foundation DMREF grant: DMR-1729166 “DMREF/GOALI: Integrated Computational Framework for Designing Dynamically Controlled Alloy -Oxide Heterostructures”.
Computational resources provided by the National Energy Research Scientific Computing Center (NERSC), supported by the Office of Science and US Department of Energy under Contract No. DE-AC02-05CH11231, are gratefully acknowledged, in addition to support from the Center for Scientific Computing from the CNSI, MRL, and NSF MRSEC (No. DMR-1720256).

%\bibliography{./references.bib}
\end{document}